\documentclass{article}
\usepackage{graphicx}
\usepackage{url}
\usepackage{color}
\usepackage{xcolor}
\usepackage{listings}
\usepackage{booktabs}
\usepackage{hyperref}
\usepackage{tabularx}

\definecolor{codegreen}{rgb}{0,0.6,0}
\definecolor{codegray}{rgb}{0.5,0.5,0.5}
\definecolor{codepurple}{rgb}{0.58,0,0.82}
\definecolor{backcolour}{rgb}{0.95,0.95,0.92}

\lstdefinestyle{mystyle}{
    backgroundcolor=\color{backcolour},   
    commentstyle=\color{codegreen},
    keywordstyle=\color{magenta},
    numberstyle=\tiny\color{codegray},
    stringstyle=\color{codepurple},
    basicstyle=\ttfamily\footnotesize,
    breakatwhitespace=false,         
    breaklines=true,                 
    captionpos=b,                    
    keepspaces=true,                 
    numbers=left,                    
    numbersep=5pt,                  
    showspaces=false,                
    showstringspaces=false,
    showtabs=false,                  
    tabsize=2
}

\title{SMOCS: A Streaming Framework for Simplified Deployment, Monitoring, and Optimization of ML Systems in Production}
\author{Armen Kasparian \and Kishansingh Rajput \and Malachi Schram \and John Vennekate}

\date{March 2026}

\begin{document}

\maketitle

\begin{abstract}
Machine learning has demonstrated significant potential for real-time monitoring, optimization, and control of scientific facilities. However, deploying and maintaining ML models in operational environments remains a substantial engineering challenge. Each facility presents unique data protocols, non-standard formats, and infrastructure constraints, forcing teams to rebuild integration pipelines for every new application. We present SMOCS (Streaming Monitoring Optimization and Control System), a Kafka-based containerized framework that addresses this challenge through three contributions: 1) a layered abstraction over Apache Kafka that separates infrastructure from application logic, 2) a three-thread agent architecture that temporally decouples data ingestion, model training, and real-time inference enabling continuous online learning from live data streams, and 3) a configuration-driven deployment model that enables domain experts to operate ML pipelines without software engineering expertise. SMOCS is facility platform-agnostic, fault-isolated by design, and horizontally scalable through Docker containerization. The framework is publicly available as open-source software at \url{https://github.com/JeffersonLab/SMOCS} with documentation at \url{https://pages.jlab.org/datascience/smocs_docs/}.
\end{abstract}

\newpage

\section{Metadata}

\begin{table}[htb]
\centering
\caption{Code metadata}
\label{tab:metadata}
\begin{tabular}{|l|p{5cm}|p{5cm}|}
\hline
C1 & Current code version & Alpha 1.0 \\
\hline
C2 & Permanent link to code/repository & \url{https://github.com/JeffersonLab/SMOCS} \\
\hline
C3 & Legal Code License & MIT \\
\hline
C4 & Code versioning system used & Git \\
\hline
C5 & Software code languages, tools, and services used & Python, Docker, Apache Kafka \\
\hline
C6 & Compilation requirements \& dependencies & Docker \\
\hline
C7 & Link to developer documentation & \url{https://pages.jlab.org/datascience/smocs_docs/}\\
\hline
C8 & Support email for questions & \url{armenk@jlab.org} \& \url{kishan@jlab.org}\\
\hline
\end{tabular}
\end{table}

\section{Introduction}

Machine learning is increasingly applied to real-time tasks in scientific facilities, from anomaly detection in particle accelerator beam lines to predictive maintenance in fusion experiments. Yet the path from a trained model to an operational deployment remains disproportionately difficult. For example, a researcher who develops an anomaly detection model may face weeks or months of integration work before that model can process live sensor data, retrain on new observations, and produce actionable diagnostics in real time. This gap between algorithmic development and operational deployment represents one of the most significant barriers to broader adoption of ML in experimental science.

These challenges arise from fundamental mismatches between ML development environments and facility infrastructure, not from any single technical obstacle. Scientific facilities typically stream data through domain-specific protocols such as Experimental Physics and Industrial Control System (EPICS) Channel Access \cite{EPICS} or Message Queuing Telemetry Transport (MQTT) \cite{mqtt_v5_2019}, using non-standard formats and timing conventions that vary across installations. Deploying an ML application on this data requires building custom protocol adapters, managing format conversions, and ensuring that the application's dependencies do not conflict with the facility's control systems. Each deployment becomes a bespoke engineering project. When another usecase and/or facility wishes to adopt the same workflow, the integration effort must largely be repeated because the original implementation was tightly coupled to the first facility's infrastructure.

Beyond initial deployment, maintaining ML systems in continuous operation introduces further challenges. Models must be retrained as system conditions evolve, yet retraining must not interrupt real-time inference. Software dependencies must be isolated so that updates to one component do not destabilize others. Failed processes must be detected and restarted without requiring human intervention, and without cascading failures that could affect facility operations.

To address these challenges we present SMOCS (Streaming Monitoring Optimization and Control System) with three principal contributions. First, it provides a layered abstraction over Apache Kafka \cite{Kreps2011KafkaA} that encapsulates all messaging infrastructure behind simple extension points, allowing developers to implement domain logic without distributed systems expertise. Second, it introduces agent architectures which enable continuous learning without service interruption. Third, it offers a configuration-driven deployment model built on Docker \cite{docker} containerization, allowing domain experts to deploy, tune, and scale ML pipelines through YAML configuration files and Docker Compose profiles without modifying application code.

SMOCS is designed to be facility-agnostic: a workflow developed for one installation can be deployed at another facility that uses the same interface by changing configuration files rather than rewriting integration code. The framework is open-source, publicly available, and includes protocol adapters for common scientific data systems, storage integrations, and reference agent implementations.

The remainder of this paper is organized as follows. Section~\ref{sec:related} reviews related work across ML deployment, stream processing, and scientific control systems. Section~\ref{sec:architecture} describes the system architecture, including the base Kafka abstractions, the three-thread agent design, and the containerization model. Section~\ref{sec:developing} demonstrates the developer experience through a worked example. Section~\ref{sec:deployment} presents built-in components and deployment workflows. Section~\ref{sec:discussion} discusses design trade-offs and limitations. Section~\ref{sec:conclusions} concludes with a summary of contributions and future directions.

\section{Related Work}
\label{sec:related}

SMOCS draws on and differs from several categories of existing tools. We organize the discussion by the primary concern each category addresses.

\subsection{ML Deployment and Lifecycle Platforms}

Platforms such as MLflow \cite{zaharia2018mlflow}, Kubeflow \cite{kubeflow}, and TensorFlow Extended (TFX) \cite{tensorflow2015-whitepaper} provide infrastructure for tracking experiments, packaging models, and orchestrating training pipelines. These tools excel at managing the model lifecycle in batch-oriented settings: a dataset is prepared, a model is trained, evaluated, and promoted to a serving endpoint. However, they are not designed for scenarios where training data arrives continuously as a stream, models must retrain periodically without interrupting inference, and the entire pipeline must operate as a persistent service rather than a scheduled job. SMOCS addresses this continuous-operation pattern through its three-thread agent architecture, where ingestion, training, and inference run concurrently as long-lived processes.

\subsection{Stream Processing Frameworks}

Apache Kafka Streams \cite{KafkaStream}, Apache Flink \cite{flink}, and Apache Spark Streaming \cite{spark} provide robust infrastructure for processing data streams at scale. These frameworks offer primitives for window ing, aggregation, and stateful transformations. A developer could build an ML pipeline on top of any of these frameworks, but doing so requires substantial effort to implement the agent like patterns that SMOCS provides as base class abstractions: coordinating concurrent training and inference threads, managing model versioning, and providing fault isolated lifecycle management for each component. SMOCS uses Kafka as its transport layer and adds agent orchestration on top.

\subsection{Scientific Control and Data Acquisition Systems}

Scientific facilities rely on control system frameworks such as EPICS for data acquisition, device control, and monitoring. These systems provide reliable, low-latency communication between sensors, actuators, and operator interfaces, but they do not include abstractions for ML workflows. Integrating an ML model typically requires building a custom bridge between the control system's data protocol and the model's runtime environment. SMOCS provides this bridge through protocol-specific producer components (ex., \textbf{EpicsKafkaProducer}, \textbf{MQTTKafkaProducer}) that translate facility data into a standardized Kafka message format, making the downstream pipeline independent of the source protocol.

\subsection{Facility Specific ML Integrations}

Several research groups have built one-off ML integrations for specific facilities. These implementations are typically tightly coupled to a single facility's infrastructure and data formats. The engineering effort required to develop each integration is substantial, and the resulting systems are difficult to port to other facilities. SMOCS aims to reduce this repeated effort by providing a reusable framework that separates facility specific protocol handling from facility agnostic ML pipeline logic.

SMOCS occupies a distinct position in this landscape: it provides ML pipeline abstractions comparable to lifecycle platforms, built on streaming infrastructure comparable to stream processing frameworks, with containerized deployment comparable to cloud-native orchestration systems, specifically designed for the continuous learning workflows common in scientific facilities.

\section{System Architecture}
\label{sec:architecture}

\begin{figure}[htb]
\centering
\includegraphics[width=\textwidth]{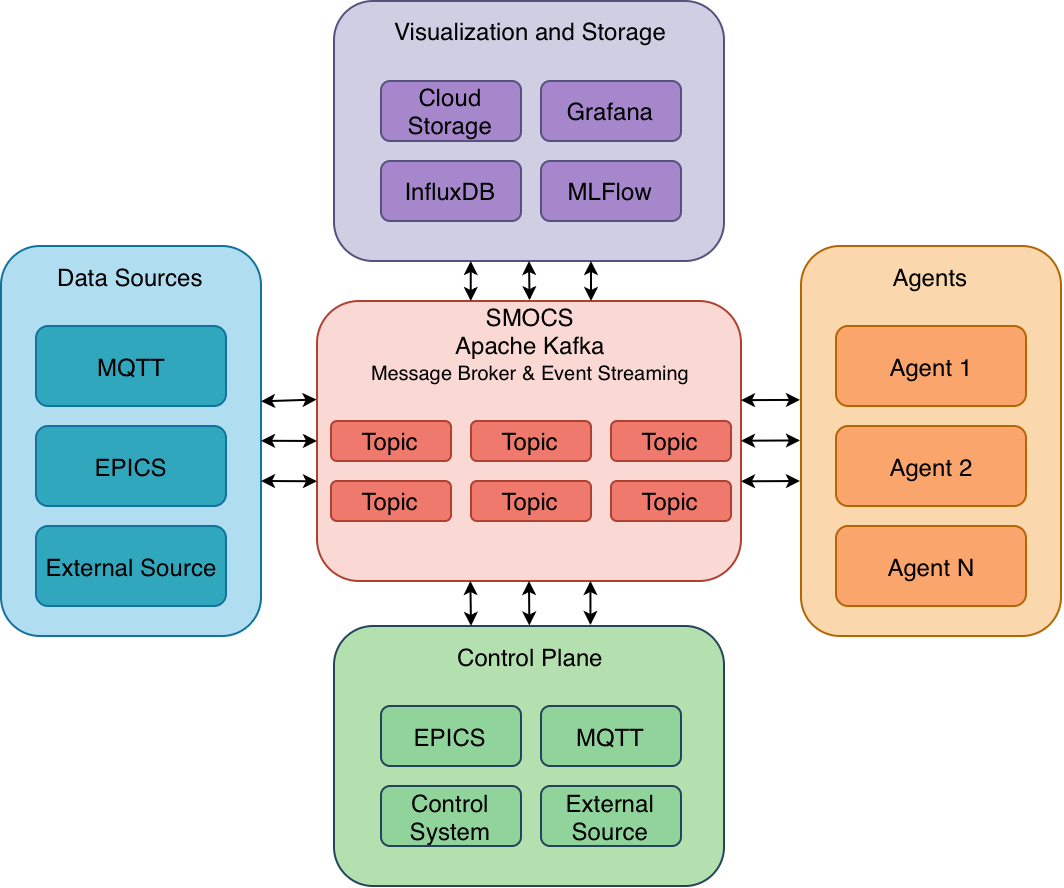}
\caption{SMOCS system architecture. \textcolor{cyan}{External data sources} connect to Kafka through protocol-specific producers (left). \textcolor{red}{The Kafka broker} manages topics and message persistence (center). \textcolor{violet}{Consumers} pull data in for visualization and storage (top). \textcolor{green}{Streaming processors} read data from SMOCS and write to the appropriate external data sources (bottom). \textcolor{orange}{Agents} subscribe to topics for transformation, and inference, and messaging (right). All components run in isolated Docker containers.}
\label{fig:architecture}
\end{figure}

SMOCS employs a layered architecture with Apache Kafka as its central messaging infrastructure. The design separates concerns across three layers: Kafka abstractions that encapsulate messaging mechanics, an agent layer that orchestrates ML workflows, and a containerization layer that provides isolation and deployment management. Figure~\ref{fig:architecture} illustrates the overall system topology.

\subsection{Architecture Overview}

SMOCS follows a hub-and-spoke topology with the Kafka broker at its center. External data sources such as facility control systems, IoT sensors, simulation environments connect through protocol-specific producer components that translate source-native formats into a standardized JSON message format within Kafka. Once data enters the Kafka broker, it becomes available to any number of downstream components: storage consumers that persist data to databases such as InfluxDB \cite{influxdb}, streaming processors that transform or enrich data in flight, and ML agents that implement continuous learning pipelines. Each component connects only to Kafka, never directly to other components, creating loose coupling that allows independent development, deployment, and scaling.

\begin{figure}[htb]
\centering
\includegraphics[width=1.0\textwidth]{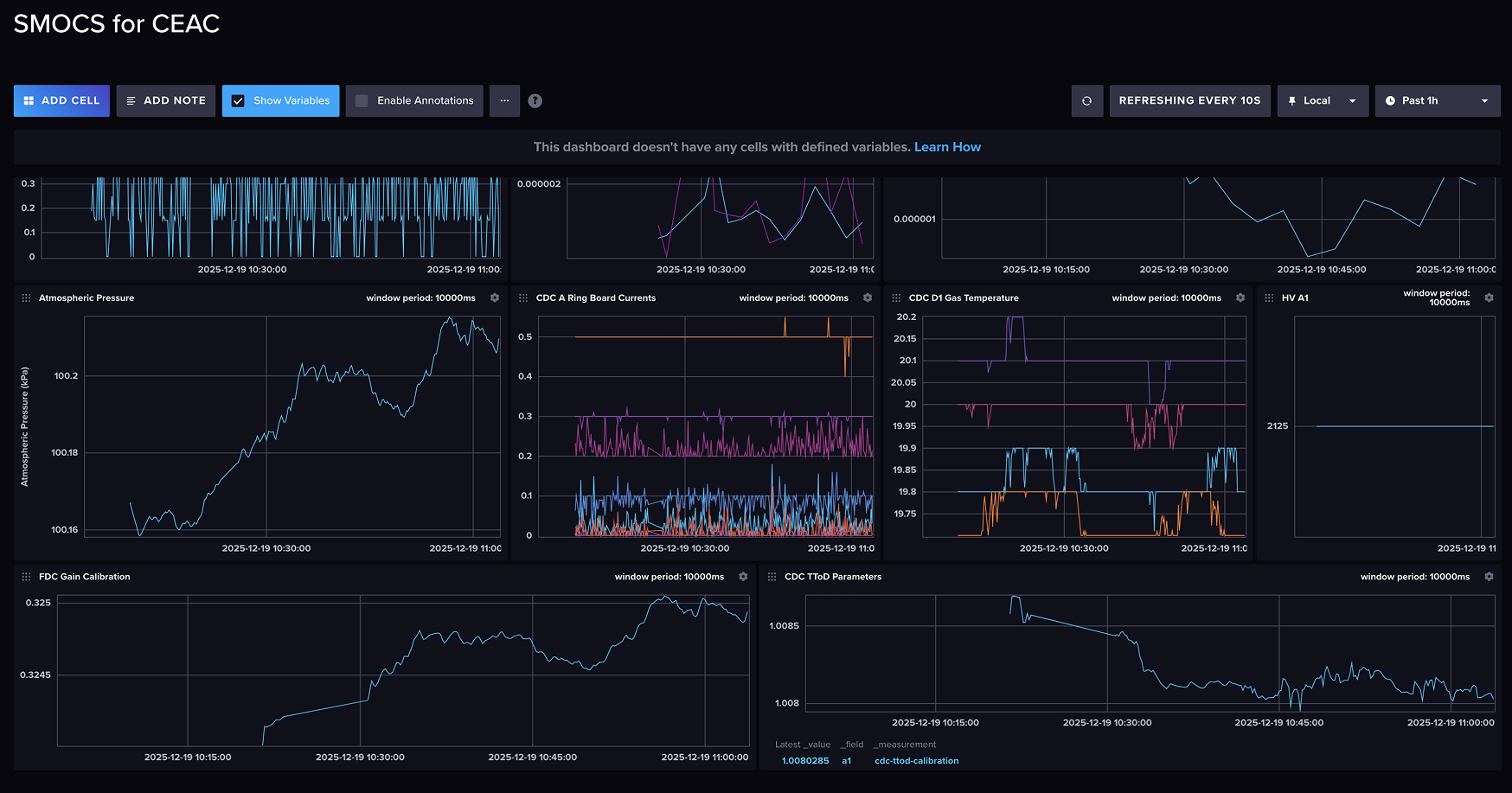}
\caption{SMOCS influxDB Dashboard being utilized to visualize live streaming data ingested from the detector halls at JLab.}
\label{fig:SMOCS_CEAC}
\end{figure}

This topology provides natural message durability and replay capability through Kafka's log-based storage. If a consumer is temporarily unavailable, messages accumulate in the broker and are delivered when the consumer reconnects. Multiple consumers can independently read from the same topic at different rates, enabling a single data stream to simultaneously feed a real-time dashboard, a long-term storage system, and an ML training pipeline. An example dashboard can be seen in fig. \ref{fig:SMOCS_CEAC} where a JLab researcher is utilizing an influxDB dashboard to visualize multiple data sources.

\subsection{Data Sources}

Data sources serve as the first step into the SMOCS system. Each data source is handled by a producer component that is tasked with connecting to an external system, reading in the native data stream, and converting it into the standardized SMOCS format which is published to our internal Kafka broker. This can be seen as the blue modules on the left of Figure \ref{fig:architecture}. The conversion is handled entirely by the producer. This enables downstream components internal to SMOCS to never have to account for variations in differing protocols. This is a powerful construct enabling reuse of SMOCS components for a multitude of facilities and systems.

For facilities that use a protocol with an existing producer implementation, connecting a new data source requires only configuration via YAML. By updating the specifics on relevant connection parameters and desired data streams users can leverage preexisting producers for their facilities. No application code needs to be written or modified leading to quick startup time when adopting SMOCS. When a facility uses a protocol that does not have a prebuilt producer available, developers can extend the \texttt{KafkaProducerBase} class to implement the necessary connection and parsing logic for their specific system. Once built, that producer becomes reusable across any facility that shares the same protocol. SMOCS currently ships with producers for EPICS and MQTT, covering two common protocols for scientific facilities and edge connected devices.

\subsection{Agents}

Agents are the primary computational components in SMOCS. At a high level, an agent consumes streaming data from Kafka, processes it according to its configured task, and publishes results back to Kafka for downstream use such as storage, visualization dashboards, and optimization/control. The scope of what an agent can accomplish is broad: agents can perform anomaly detection and system monitoring, predictive maintenance, reinforcement learning for optimization and control, simple threshold or statistical checks, or data transformation and feature engineering. SMOCS imposes no constraints on the complexity of the task, and the same agent architecture supports use cases ranging from flagging a sensor value that exceeds a fixed bound to training a deep neural network for real-time diagnostics. For those interested in the internal design of agents or in developing custom agent implementations, Section~\ref{sec:developing} provides a detailed walkthrough.

Deploying an agent is an operational task rather than a development task when a suitable agent type already exists. An operator selects an agent type and configures its behavior via a YAML file defining parameters such as monitored data channels, threshold values, and publication endpoints, then launches it as a Docker container. Any number of agents can run simultaneously in isolated containers. Multiple agents can pull the same or different data streams; Kafka manages all message delivery independently. Due to this isolation, adding or removing agents does not affect other system components; thus, the number of concurrent agents can be scaled up or down based solely on available hardware resources.

\subsection{Control Plane}

The control plane represents the return path from the SMOCS ecosystem back to external facility systems. When an output that results in an action or signal that needs to be fed back to the physical system, such as adjusting a setpoint, issuing a command, or updating a parameter, it publishes that output to a Kafka topic like any other message. A control plane component subscribes to the relevant topic, reads the incoming messages, and translates them into the facility's native protocol for delivery to the target system. In this sense, the control plane functions as the conceptual mirror of a data source producer: where producers bring external data into SMOCS through Kafka, control plane components carry processed results back out.

SMOCS provides base class abstractions for building control plane components, and specific implementations are developed by the user to match the requirements of their facility's control and safety infrastructure.

\subsection{Visualization and Storage}

Visualization and storage components consume data from Kafka topics and direct it to external systems for persistence, monitoring, and analysis. SMOCS ships with built-in support for InfluxDB as a time-series data store and Grafana \cite{grafana} for real-time dashboarding. Together, these components allow operators to visualize raw sensor data, agent outputs, anomaly scores, and training metrics on unified dashboards without writing any custom code. For deployments that use these built-in integrations, connecting them to the appropriate Kafka topics is a configuration-only task handled through the Docker Compose files.

Because SMOCS uses Apache Kafka as its central broker, the storage layer is naturally extensible. Any system that can consume from Kafka can be connected as a storage or visualization target. Users can write custom consumers to route data to cloud storage services, relational databases, logging platforms, or any other backend that fits their facility's infrastructure. Existing Kafka ecosystem tooling and connectors can also be leveraged. This flexibility ensures that SMOCS does not impose constraints on where data ultimately resides or how it is presented to operators.

\subsection{Containerization and Deployment}

SMOCS achieves environmental isolation and deployment flexibility through Docker containerization. Each component runs in its own container with explicitly declared dependencies and resource constraints. This approach solves several challenges simultaneously: dependency conflicts between components are eliminated by construction, containers can be deployed on any host that has Docker regardless of its native software environment, and failed containers can be restarted independently without affecting the rest of the system.

Deployment is driven by YAML configuration files and Docker Compose profiles. Operators select which components to deploy by specifying profiles, configure agent behavior through the central YAML configuration file, and start the system with standard Docker Compose commands. Multiple agents can run concurrently, each in its own container with an independent database, processing different data streams without interference. This enables horizontal scaling: a facility can deploy dozens of agents monitoring different subsystems simultaneously, adding or removing agents without restarting existing ones.

The configuration-driven approach separates operational concerns from application code. An operator can change which sensor channels an agent monitors, adjust training hyperparameters, or redirect output topics by editing a configuration file and restarting the relevant container. No software development expertise is required for these operational adjustments.

\section{Implementation and Deployment}
\label{sec:deployment}

This section demonstrates the practical application of the SMOCS framework through two distinct deployment scenarios. The first workflow utilizes an environment from Gymnasium to showcase how SMOCS can streamline the development and monitoring of reinforcement learning agents. The second workflow transitions to a industrial application: multi-agent beam monitoring using data from the CEBAF accelerator. Together, these examples illustrate SMOCS's ability to handle production environments by leveraging a decoupled, Kafka-based architecture that ensures scalability and modularity.

\subsection{Workflow: Gymnasium w/ Reinforcement Learning}

The first demonstration workflow uses SMOCS to train a reinforcement learning agent on a continuous control task. The workflow illustrates how the streaming processor and agent abstractions compose into a training system where the environment, policy, and monitoring infrastructure communicate entirely through Kafka.

The workflow centers on the Pendulum-v1 environment from Gymnasium, a standard continuous control benchmark in which an agent must learn to swing up and balance an inverted pendulum by applying torque. The SMOCS deployment consists of five containerized components: the \texttt{KafkaGymWrapper} streaming processor running the Pendulum-v1 environment, a Twin Delayed Deep Deterministic Policy Gradient (TD3)\cite{td3} reinforcement learning agent built on Jefferson Lab's Scientific Optimization and Control Toolkit (SOCT) \cite{SOCT}, an autoencoder anomaly detection agent monitoring the environment's state space, an InfluxDB consumer persisting all metrics for visualization, and the Kafka broker coordinating all message passing.

\begin{figure}[htb]
\centering
\includegraphics[width=0.6\textwidth]{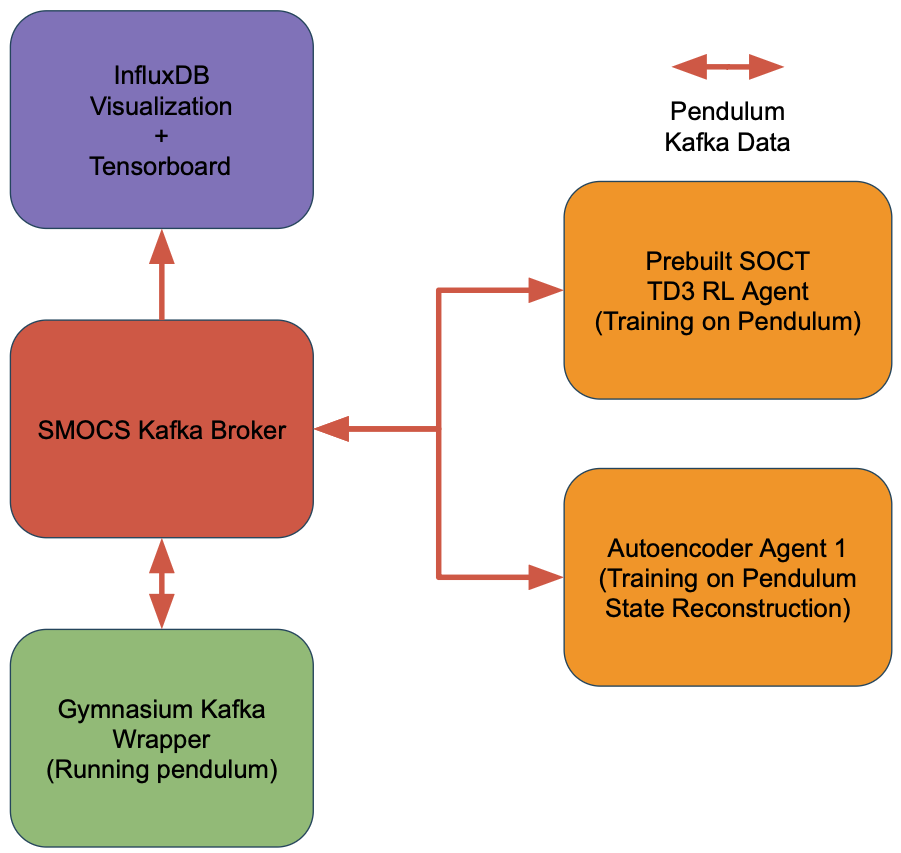}
\caption{Reinforcement learning training workflow. The Gymnasium wrapper publishes environment states to Kafka. The RL agent's inference thread generates actions, which flow back to the environment through Kafka. SARSA transition tuples are simultaneously routed to the agent's ingest thread for experience replay storage. The autoencoder agent independently monitors the same state stream for anomaly detection. All metrics flow to InfluxDB for visualization.}
\label{fig:rl_workflow}
\end{figure}

Figure~\ref{fig:rl_workflow} illustrates the data flow through this system.

The data flow operates in a closed loop. The Gymnasium wrapper executes environment steps and publishes three types of messages to separate Kafka topics: state observations for the inference thread, complete SARSA (state, action, reward, next state, done) transition tuples for the training thread, and decomposed channel data for monitoring. The RL agent's inference thread consumes state messages and publishes actions back to Kafka, where the Gymnasium wrapper receives them and executes the corresponding environment step. The RL agent's ingest thread simultaneously consumes the SARSA tuples and stores them in its experience replay buffer. Once sufficient experience has accumulated (2,500 samples in the default configuration), the training thread begins updating the TD3 policy and critic networks on each new experience.

The autoencoder agent operates independently on the same state stream, learning to reconstruct the pendulum's state (cosine and sine of the angle, angular velocity). Because the RL agent's exploration produces state distributions that evolve over time the autoencoder provides a complementary view of the learning process. Reconstruction error increases during state distribution shifts, offering an indirect signal that the RL policy is changing the system's behavior.

The entire workflow is deployed by setting a single environment variable (\texttt{COMPOSE\_PROFILES=gymnasium,rl1,autoencoder1}) and running  \textbf{docker compose up}. In practice, the system reaches steady-state operation: the RL agent's episode rewards improve from approximately $-1500$ (random policy) toward $-200$ (near-optimal control), the autoencoder trains its first model after accumulating sufficient data, and anomaly detection begins flagging state deviations. Training progress is monitored through InfluxDB dashboards.

\subsection{Workflow: Multi-Agent Beam Monitoring}

The second workflow demonstrates SMOCS's ability to rapidly deploy and reconfigure multiple agents against a single data source, illustrating the operational flexibility that distinguishes the framework from one-off implementations.

\begin{figure}[htb]
\centering
\includegraphics[width=0.95\textwidth]{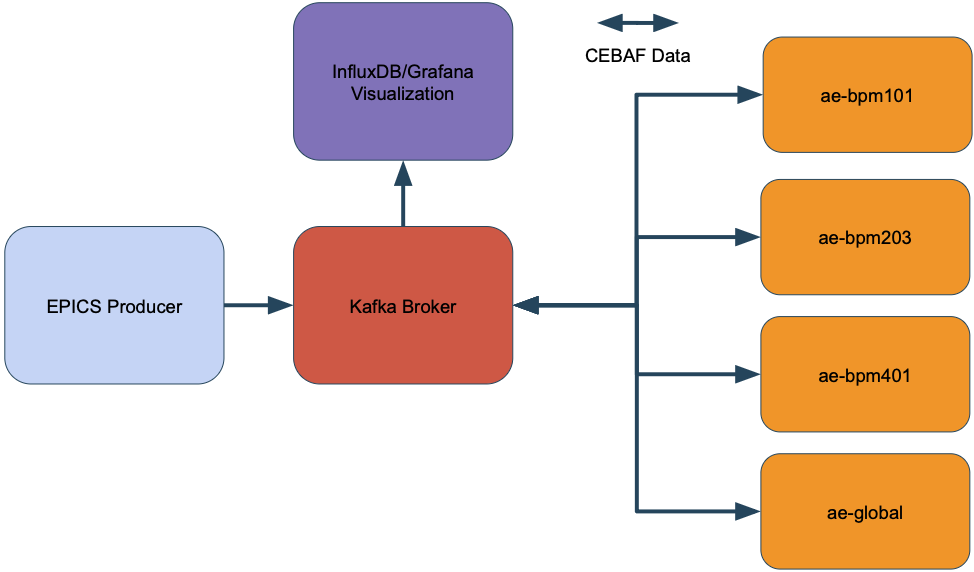}
\caption{Multi-agent beam monitoring topology. An EPICS producer publishes CEBAF BPM data to a single Kafka topic. Four autoencoder agents consume the same topic but filter different channel subsets based on their YAML configuration. The InfluxDB consumer aggregates anomaly outputs from all agents for visualization.}
\label{fig:multi_agent}
\end{figure}

The scenario uses live beam position monitor (BPM) data from the Continuous Electron Beam Accelerator Facility (CEBAF) at Jefferson Lab. An EPICS producer ingests X position, Y position, and phase measurements from three BPMs: IPMK101, IPMK203, and IPMK401. These BPMs produce a stream of nine sensor channels on a single Kafka topic. Four autoencoder agents are deployed simultaneously against this stream, each configured to monitor a different set of variables of the data as seen in figure~\ref{fig:multi_agent}. 

\begin{table}[htb]
\centering
\caption{Four-agent deployment monitoring different aspects of CEBAF beam position data. All agents consume the same Kafka topic but filter different channel subsets through configuration.}
\label{tab:multi_agent}
\begin{tabular}{@{}llp{5.5cm}@{}}
\toprule
\textbf{Agent} & \textbf{Channels} & \textbf{Monitoring Objective} \\
\midrule
\texttt{ae-bpm101} & IPMK101 X, Y, phase & Single beam health at one location \\
\texttt{ae-bpm203} & IPMK203 X, Y, phase & Single beam health at a second location \\
\texttt{ae-bpm401} & IPMK401 X, Y, phase & Single beam health at a third location \\
\texttt{ae-global} & All 9 channels & Cross-BPM correlation and system-wide beam stability \\
\bottomrule
\end{tabular}
\end{table}

Table~\ref{tab:multi_agent} illustrates the deployment topology.

The critical point is that none of these agents required any code changes to deploy. Each is an instance of the same autoencoder agent class, differentiated entirely by a YAML configuration block that specifies which channels to monitor, the normalization bounds for each channel, and the output topic name. For example, adding a new agent that monitors only the X positions across all three BPMs to detect horizontal orbit drift, or one that monitors only phase channels to flag RF timing issues only requires writing a new configuration block and adding a Docker Compose entry. The new agent is deployed by appending its profile name to the environment variable and running \texttt{docker compose up}. No existing agents are restarted or affected.

This ability to rapidly spool agents up and down based on operational needs is a direct design principle of SMOCS's architecture. Because each agent runs in an isolated container with its own MySQL database, model filesystem, and Kafka consumer group, deploying or removing an agent has no side effects on the rest of the system. An operator investigating a suspected issue with horizontal beam steering could spin up a specialized agent monitoring only X positions, observe its behavior for a shift, and spool it down all without touching the persistent monitoring agents that provide continuous coverage. Similarly, during a machine study period with unusual beam configurations, an operator might temporarily deploy agents with relaxed anomaly thresholds or different channel groupings, then revert to the standard configuration when normal operations resume. 

The InfluxDB consumer aggregates anomaly outputs from all four agents into a single database, enabling operators to view raw sensor data and anomaly scores on a unified InfluxDB or Grafana dashboard.

\section{Developing with SMOCS}
\label{sec:developing}

SMOCS provides a structured framework for developing custom components. This section describes the message format standard, walks through a representative agent implementation, and summarizes the development patterns available to developers. Complete implementation examples and API documentation are available in the SMOCS documentation \cite{Kasparian}.

\subsection{Message Format Standard}

All data flowing through SMOCS adheres to a standardized JSON format that ensures interoperability between components. Every Kafka message contains two required fields:

\begin{lstlisting}[language=Python]
{
  "timestamp": <unix_timestamp_or_iso_string>,
  "channels": {
    "sensor_name_1": value_1,
    "sensor_name_2": value_2
  }
}
\end{lstlisting}

The \texttt{timestamp} field accepts Unix epoch values (integers or floats), ISO-format strings, or null values for timestamp-agnostic data. The \texttt{channels} dictionary contains arbitrary key-value pairs representing sensor measurements, actuator commands, or derived quantities. This separation of temporal information from payload data enables downstream components to process messages uniformly regardless of their origin. SMOCS base classes automatically validate message compliance, allowing developers to focus on domain logic rather than format checking.

\subsection{Base Kafka Abstractions}

The foundation of SMOCS consists of three abstract base classes that encapsulate all Kafka-specific functionality. These abstractions establish a clear contract: the framework manages connections, polling, error recovery, serialization, and offset management, while developers implement only the application logic specific to their use case.
Table~\ref{tab:abstractions} summarizes the three base classes and their responsibilities.

\begin{table}[htb]
\centering
\caption{SMOCS base abstractions. Each class encapsulates Kafka infrastructure and defines a minimal interface for developer extension.}
\label{tab:abstractions}
\footnotesize
\begin{tabular}{@{}p{3.2cm}p{4.2cm}p{3.8cm}@{}}
\toprule
\textbf{Base Class} & \textbf{Encapsulates} & \textbf{Developer Implements} \\
\midrule
\texttt{KafkaConsumer\-Base} & Subscription, polling, error recovery & \texttt{process\_message()} \\
\addlinespace
\texttt{KafkaProducer\-Base} & Publishing, topic management, serialization & Protocol-specific ingestion \\
\addlinespace
\texttt{KafkaStreaming\-ProcessBase} & Bidirectional Kafka I/O & \texttt{process\_message()}\\
\bottomrule
\end{tabular}
\end{table}

The \texttt{KafkaConsumerBase} establishes the pattern for all data consumption. It manages topic subscription (supporting both explicit topic lists and regex patterns), implements a standardized polling loop, and provides error handling and resource cleanup. Subclasses implement a single \texttt{process\_message()} method containing their application logic.

The \texttt{KafkaProducerBase} handles reliable message publishing with automatic topic creation, message serialization, and topic name sanitization (for example, converting MQTT-style hierarchical topic names to valid Kafka identifiers). Data source integrations extend this class and implement their specific protocol handling.

The \texttt{KafkaStreamingProcessBase} combines consumption and production into a single component. It inherits from \texttt{KafkaConsumerBase} for input handling and composes a \texttt{KafkaProducerBase} instance for output publishing. The developer's \texttt{process\_message()} method returns both a success flag and a list of output destination-message pairs, enabling transformations, filtering, enrichment, and fan-out patterns through a single interface.

A deliberate design choice across all base classes is single-threaded execution. Each Kafka abstraction processes messages sequentially within a single thread. This simplifies error handling and recovery, makes system behavior deterministic and easier to reason about during debugging, and eliminates concurrency-related failure modes within each component. Parallelism is achieved at the component level through Kafka's consumer group protocol as multiple instances of the same consumer can share a topic's messages.

\subsection{Three-Thread Agent Architecture}

\begin{figure}[htb]
\centering
\includegraphics[width=0.95\textwidth]{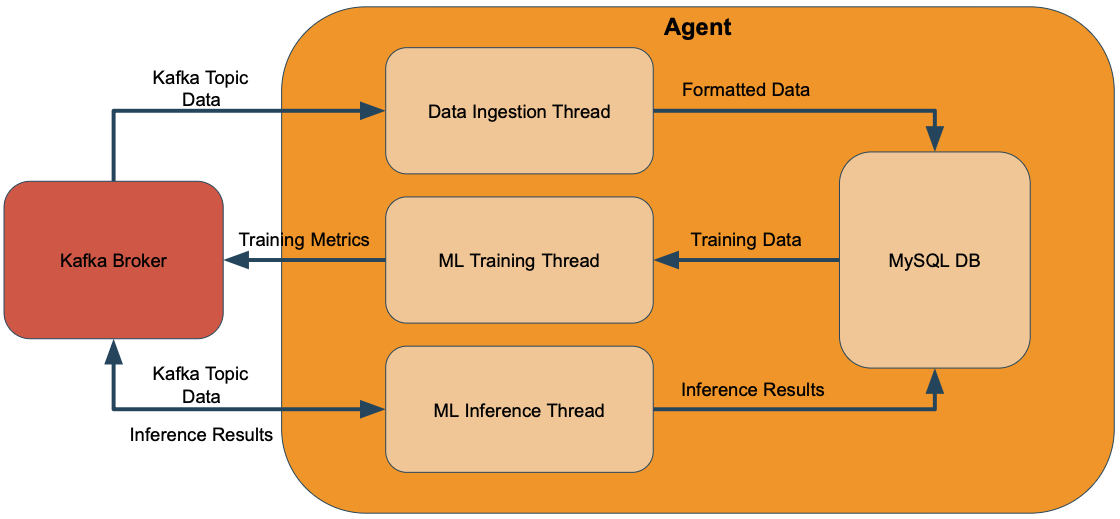}
\caption{Three-thread agent architecture. The Data Ingest Thread consumes sensor data from Kafka at message arrival rate and writes to MySQL. The Training Thread periodically queries accumulated data, trains models, and writes versioned artifacts to the filesystem. The Inference Thread consumes live data, loads the latest model, and publishes predictions to Kafka output topics. Threads coordinate through the shared database and filesystem without direct communication.}
\label{fig:agent}
\end{figure}

SMOCS agents extend the base abstractions into a complete ML pipeline through the coordination of three specialized threads, each optimized for a distinct operational tempo. Figure~\ref{fig:agent} illustrates this architecture. The three threads run asynchronously and communicate internally via an agent specific MySQL database. This allows the agent to communicate internally and pass data between the threads without the need of complex message passing structures or locks.

The \textbf{Data Ingest Thread} inherits from \texttt{KafkaConsumerBase} and operates at the sensor sampling rate. It consumes streaming sensor data from configured Kafka topics, applies agent-specific parsing and validation, and stores structured readings to the agent's internal MySQL database. This thread creates a continuously growing dataset that serves as the foundation for model training.

The \textbf{Training Thread} inherits from \texttt{KafkaProducerBase} and operates on a periodic schedule rather than in response to incoming messages. At configurable intervals, it queries the database for accumulated training data, evaluates whether sufficient new data has arrived since the last training cycle, and when configurable conditions are met executes the full training workflow: data retrieval, preprocessing, model training, evaluation, and versioned model persistence. Training outcomes and metrics are published to Kafka topics for system-wide monitoring.

The \textbf{Inference Thread} inherits from \texttt{KafkaStreamingProcessBase} and operates at message arrival rate. It consumes real-time sensor data, processes each reading through the latest trained model, and publishes inference results (predictions, anomaly scores, confidence measures, etc depending on the model) to output topics. The thread implements dynamic model loading, automatically detecting and incorporating new model versions as they become available from the training thread.

The key architectural insight is \textit{temporal decoupling}. The three threads operate on fundamentally different time scales. Data ingestion (potentially hundreds of messages per second), periodic training (minutes to hours), and message-rate inference (continuous) are all happening potentially simultaneously and asynchronously yet they coordinate without blocking one another. This coordination occurs through two shared resources rather than direct inter-thread communication. The first is the MySQL database which mediates data flow internal to the agent: the ingest thread writes records, and the training thread reads them. The second is the filesystem which mediates model handoff: the training thread writes versioned model artifacts and updates a metadata file, and the inference thread polls this metadata to detect new versions. Atomic file operations ensure the inference thread never loads a partially written model.

This design provides fault isolation at the thread level. If the inference thread crashes due to a malformed input, the ingest and training threads continue operating, preserving data collection and model development. The \texttt{AgentBase} class continuously monitors thread health and automatically restarts failed components without affecting the others. This resilience is essential for systems that must operate continuously in facility environments where unattended recovery is expected.

\subsection{Built-in Components}

SMOCS includes several predeveloped components that address common integration needs. The \texttt{MQTTKafkaProducer} bridges MQTT brokers into the Kafka ecosystem, translating MQTT message formats and topic hierarchies into the standardized SMOCS JSON format. The \texttt{EpicsKafkaProducer} provides equivalent functionality for EPICS Channel Access, enabling integration with accelerator and experimental physics control systems. Both producers handle connection management, reconnection logic, and configurable data parsing through the central configuration file.

On the consumption side, the \texttt{InfluxDBConsumer} subscribes to Kafka topics via regex patterns, selectively processes messages based on type markers, and persists time-series data to InfluxDB for long-term storage and visualization. This component demonstrates how SMOCS consumers can implement intelligent filtering and routing logic while maintaining the simple single-threaded processing model.

The \texttt{KafkaGymWrapper} exemplifies the streaming processor pattern by bridging SMOCS with Gymnasium reinforcement learning environments \cite{towers2025gymnasiumstandardinterfacereinforcement}. It consumes action commands from Kafka, executes them in a Gymnasium environment, and publishes complete state-transition tuples (state, action, reward, next state) back to Kafka. The wrapper supports both blocking and non-blocking execution modes, demonstrating how the base architecture accommodates different operational patterns for RL training workflows.

\subsection{Worked Example: Autoencoder Anomaly Detection Agent}

To illustrate the developer experience, we walk through the construction of an autoencoder-based anomaly detection agent. This agent monitors a set of sensor channels, learns the normal operating distribution through an autoencoder, and flags readings that deviate significantly from learned patterns. Building this agent requires implementing eight methods across three thread classes, summarized in Table~\ref{tab:agent_methods}.

\begin{table}[htb]
\centering
\caption{Methods required to implement a complete anomaly detection agent. Each method encapsulates one concern; the framework handles all orchestration.}
\label{tab:agent_methods}
\begin{tabularx}{\columnwidth}{@{}llX@{}}
\toprule
\textbf{Thread} & \textbf{Method} & \textbf{Responsibility} \\
\midrule
Data Ingest & \texttt{store\_message()} & Parse sensor data from Kafka message, store to MySQL \\
\midrule
Training & \texttt{get\_training\_data()} & Query database, check for sufficient new data \\
         & \texttt{train\_model()} & Train autoencoder on accumulated data \\
         & \texttt{eval\_model()} & Compute reconstruction error distribution, set anomaly threshold \\
         & \texttt{save\_model()} & Persist model with version metadata \\
\midrule
Inference & \texttt{load\_model()} & Detect and load latest model version \\
          & \texttt{parse\_inference\_request()} & Extract sensor values from incoming message \\
          & \texttt{perform\_inference()} & Reconstruct input, compare error to threshold, flag anomalies \\
\bottomrule
\end{tabularx}
\end{table}

The data flow through the agent proceeds as follows. The ingest thread receives sensor messages from Kafka, extracts the configured channels (ex., temperature, pressure, flow rate), and stores timestamped readings to MySQL. As data accumulates, the training thread periodically queries the database, constructs training samples, trains an autoencoder to minimize reconstruction error, and computes an anomaly threshold as appropriate on an evaluation set. The trained model and threshold are saved to the filesystem with incrementing version numbers. Meanwhile, the inference thread maintains a sliding window buffer of recent sensor readings. For each new message, it constructs the current window, passes it through the autoencoder, computes the reconstruction error, and compares it against the learned threshold. Results are published to a configured Kafka output topic.

Listing~\ref{lst:inference} shows the \texttt{perform\_inference()} method, which represents the core diagnostic logic. This is the most algorithmically substantive of the eight methods; the others are comparably concise.

\begin{lstlisting}[language=Python, caption={Inference method for the autoencoder anomaly detection agent. The method maintains a sliding window buffer and compares reconstruction error against a learned threshold.}, label={lst:inference}]
def perform_inference(self, inference_request):
    self.recent_data.append(inference_request['sensor_values'])
    
    if len(self.recent_data) < self.window_size:
        return None  # Insufficient data for full window
    
    window = np.array(self.recent_data[-self.window_size:])
    flattened = window.flatten().reshape(1, -1)
    reconstruction = self.model.predict(flattened, verbose=0)
    error = float(np.mean((flattened - reconstruction) ** 2))
    
    return {
        'error_score': error,
        'is_anomaly': error > self.anomaly_threshold,
        'threshold': self.anomaly_threshold,
        'status': 'success'
    }
\end{lstlisting}

The agent is configured entirely through a YAML file that specifies which threads to enable, which sensor channels to monitor, model architecture parameters, and Kafka topic mappings:

\begin{lstlisting}[language=Python, caption={Agent configuration. All operational parameters are externalized, enabling rapid experimentation without code changes.}, label={lst:config}]
autoencoder1:
  enabled_threads: ['ingest', 'training', 'inference']
  model_input:
    channels: [temperature, pressure, flow_rate]
    bounds: [[0.0, 100.0], [0.0, 50.0], [0.0, 10.0]]
  window_size: 50
  encoder_dims: [32, 16]
  learning_rate: 0.001
  min_training_samples: 10000
  batch_size: 32
  epochs: 50
  kafka_topics:
    input: "sensor-data"
    output: "autoencoder-anomalies"
    training_output: "autoencoder-training"
\end{lstlisting}

Deployment requires a Docker Compose entry that maps the agent to a container and associates it with a deployment profile:

\begin{lstlisting}[language=bash, caption={Docker Compose configuration for the agent. The \texttt{extends} directive inherits base agent infrastructure.}, label={lst:compose}]
autoencoder-agent1:
  extends: agent
  profiles: ["autoencoder1"]
  environment:
    AGENT_TYPE: autoencoder
    AGENT_CONFIG: autoencoder1
  volumes:
    - mysql-data-autoencoder1:/var/lib/mysql
\end{lstlisting}

The agent is then started with \texttt{docker compose up autoencoder-agent1}. Multiple agents with different configurations can be deployed simultaneously by adding entries to the YAML file and Docker Compose configuration, each running in isolation with independent database storage.

\subsection{Development Patterns Summary}

SMOCS supports several categories of custom components, each requiring a minimal implementation surface:

\textit{Streaming processors} extend \texttt{KafkaStreamingProcessBase} and implement one method, \texttt{process\_message()}, which returns a success flag and a list of output topic-message pairs. The base class handles all Kafka mechanics. Processors are stateless and horizontally scalable through Kafka consumer groups.

\textit{Agent components} are organized into three thread types. Ingest threads implement \texttt{store\_message()} for data persistence. Training threads implement four methods covering the training lifecycle: data retrieval, model training, evaluation, and persistence. Inference threads implement three methods for model loading, request parsing, and prediction. The \texttt{AgentBase} class orchestrates thread lifecycle, health monitoring, and automatic restart.

\textit{Configuration} drives all operational parameters through YAML files, enabling concurrent agents with different settings, rapid hyperparameter iteration, and deployment by operators without software development backgrounds.

This separation of concerns allows the same framework to support use cases ranging from simple unit conversion processors to sophisticated multi-agent anomaly detection systems, with the infrastructure layer handled uniformly by the SMOCS framework.

\section{Discussion}
\label{sec:discussion}

\subsection{Design Trade-offs}

Several design decisions in SMOCS were made in an attempt to balance complexity with ease of use. The single-threaded execution model within each Kafka abstraction prioritizes determinism and debuggability over maximum throughput. For most scientific facility workloads, where sensor data rates range from tens, hundreds to thousands of messages per second, this is not a limiting factor and the operational simplicity is a significant advantage. For workloads requiring higher throughput, SMOCS supports horizontal scaling by deploying multiple instances of the same component across Kafka partitions.

The use of MySQL as the coordination mechanism between agent threads introduces a dependency that simpler alternatives (such as shared memory or file-based communication) would avoid. However, MySQL provides durable storage that survives container restarts, structured querying that supports the training thread's need to sample batches from accumulated data in model specific formats, and a tested operational model. The \texttt{DBManager} abstraction isolates threads from SQL details, and the per agent database architecture prevents cross agent interference and potential lag.

The configuration driven deployment model trades some flexibility for operational accessibility. Complex conditional logic or dynamic behavior that depends on runtime state cannot be expressed in static YAML files. For such cases, developers implement custom logic in their thread methods. The configuration layer addresses the more common case where an operator needs to adjust channels, thresholds, or hyperparameters without modifying code.

\subsection{Future Work \& Limitations}

Several directions for future development are planned. Integration with large language models through the Model Context Protocol (MCP) would enable natural-language interfaces for agent deployment, monitoring, and diagnostics, lowering the barrier to entry for operators unfamiliar with configuration files. This harness would be built as an MCP enabling users to provide tooling to LLMs to interface with SMOCS. Additional custom producers for standard control planes can be developed to expand SMOCS's facility coverage. The development of SMOCS as a framework for online continual learning is also on the docket. Finally, community contributions of agent implementations for common facility workflows, beam loss monitoring, magnet quench detection, vacuum system diagnostics, alarm monitoring, and general system optimization and automation would accelerate adoption across the scientific community.

SMOCS has not yet been benchmarked under very high message rates (tens of thousands of messages per second). While Kafka itself scales to millions of messages per second, the single-threaded processing model and JSON serialization in SMOCS components may introduce bottlenecks at extreme throughput levels. Characterizing these limits and exploring binary serialization formats for high-rate applications is an area for future work. The potential for switching to a protobuf standard exists and is a potential future implementation that may be explored as needed.

The current agent architecture assumes a single model per agent. Ensemble methods or multi-model pipelines would require either multiple agents or extensions to the training and inference thread interfaces. Similarly, the three-thread design is oriented toward supervised and unsupervised learning workflows; reinforcement learning agents that require tighter coupling between inference and environment interaction may benefit from a modified thread architecture, as partially demonstrated by the Gymnasium wrapper.

\section{Conclusions}
\label{sec:conclusions}

We have presented SMOCS, a Kafka-based containerized framework for deploying machine learning workflows against streaming data in scientific facilities. The framework's principal contributions are a layered Kafka abstraction that separates infrastructure from application logic, a three-thread agent architecture that temporally decouples ingestion, training, and inference for continuous learning without service interruption, and a configuration driven deployment model that enables facility operators to manage ML pipelines without software engineering expertise.

SMOCS addresses a systemic challenge in scientific computing: the repeated, facility specific engineering effort required to bridge the gap between ML research and operational deployment. By providing reusable abstractions for the common patterns of streaming ML workflows, protocol adaptation, data persistence, model lifecycle management, and real-time inference, SMOCS aims to shift the effort from infrastructure integration toward the domain-specific algorithms that deliver scientific value.

The framework is open-source and publicly available at \url{https://github.com/JeffersonLab/SMOCS}, with documentation at \url{https://pages.jlab.org/datascience/smocs_docs/}. We welcome contributions from the scientific computing community.

\bibliographystyle{plain}
\bibliography{references}

@article{EPICS,
  author    = {Dalesio, L. R. and Hill, J. O. and Kraimer, M. and
               Lewis, S. and Murray, D. and Hunt, S. and
               Watson, W. and Clausen, M. and Dalesio, J.},
  title     = {The Experimental Physics and Industrial Control System
               Architecture: Past, Present, and Future},
  journal   = {Nuclear Instruments and Methods in Physics Research
               Section A: Accelerators, Spectrometers, Detectors
               and Associated Equipment},
  volume    = {352},
  number    = {1--2},
  pages     = {179--184},
  year      = {1994},
  doi       = {10.1016/0168-9002(94)91493-1},
}

@techreport{mqtt_v5_2019,
  title        = {{MQTT} Version 5.0},
  author       = {Banks, Andrew and Briggs, Ed and Borgendale, Ken and Gupta, Rahul},
  year         = {2019},
  month        = mar,
  type         = {{OASIS} Standard},
  url          = {https://docs.oasis-open.org/mqtt/mqtt/v5.0/os/mqtt-v5.0-os.html},
  institution  = {OASIS}
}

@article{zaharia2018mlflow,
  title     = {Accelerating the Machine Learning Lifecycle with {MLflow}},
  author    = {Zaharia, Matei A. and Chen, Andrew and Davidson, Aaron and Ghodsi, Ali and Hong, Sue Ann and Konwinski, Andy and Murching, Siddharth and Nykodym, Tomas and Ogilvie, Paul and Parkhe, Mani and Xie, Fen and Zumar, Corey},
  journal   = {IEEE Data Eng. Bull.},
  volume    = {41},
  pages     = {39--45},
  year      = {2018},
  url       = {https://api.semanticscholar.org/CorpusID:83459546}
}

@misc{kubeflow,
  title        = {Kubeflow: The Machine Learning Toolkit for Kubernetes},
  author       = {{Kubeflow Contributors}},
  year         = {2018},
  howpublished = {\url{https://www.kubeflow.org}},
  note         = {GitHub: \url{https://github.com/kubeflow/kubeflow}}
}

@misc{tensorflow2015-whitepaper,
title={ {TensorFlow}: Large-Scale Machine Learning on Heterogeneous Systems},
url={https://www.tensorflow.org/},
note={Software available from tensorflow.org},
author={
    Mart\'{i}n~Abadi and
    Ashish~Agarwal and
    Paul~Barham and
    Eugene~Brevdo and
    Zhifeng~Chen and
    Craig~Citro and
    Greg~S.~Corrado and
    Andy~Davis and
    Jeffrey~Dean and
    Matthieu~Devin and
    Sanjay~Ghemawat and
    Ian~Goodfellow and
    Andrew~Harp and
    Geoffrey~Irving and
    Michael~Isard and
    Yangqing Jia and
    Rafal~Jozefowicz and
    Lukasz~Kaiser and
    Manjunath~Kudlur and
    Josh~Levenberg and
    Dandelion~Man\'{e} and
    Rajat~Monga and
    Sherry~Moore and
    Derek~Murray and
    Chris~Olah and
    Mike~Schuster and
    Jonathon~Shlens and
    Benoit~Steiner and
    Ilya~Sutskever and
    Kunal~Talwar and
    Paul~Tucker and
    Vincent~Vanhoucke and
    Vijay~Vasudevan and
    Fernanda~Vi\'{e}gas and
    Oriol~Vinyals and
    Pete~Warden and
    Martin~Wattenberg and
    Martin~Wicke and
    Yuan~Yu and
    Xiaoqiang~Zheng},
  year={2015},
}

@INPROCEEDINGS{flink,
  author={Katsifodimos, Asterios and Schelter, Sebastian},
  booktitle={2016 IEEE International Conference on Cloud Engineering Workshop (IC2EW)}, 
  title={Apache Flink: Stream Analytics at Scale}, 
  year={2016},
  volume={},
  number={},
  pages={193-193},
  doi={10.1109/IC2EW.2016.56}}

@inproceedings{KafkaStream,
author = {Wang, Guozhang and Chen, Lei and Dikshit, Ayusman and Gustafson, Jason and Chen, Boyang and Sax, Matthias J. and Roesler, John and Blee-Goldman, Sophie and Cadonna, Bruno and Mehta, Apurva and Madan, Varun and Rao, Jun},
title = {Consistency and Completeness: Rethinking Distributed Stream Processing in Apache Kafka},
year = {2021},
isbn = {9781450383431},
publisher = {Association for Computing Machinery},
address = {New York, NY, USA},
url = {https://doi.org/10.1145/3448016.3457556},
doi = {10.1145/3448016.3457556},
booktitle = {Proceedings of the 2021 International Conference on Management of Data},
pages = {2602–2613},
numpages = {12},
location = {Virtual Event, China},
series = {SIGMOD '21}
}

@article{spark,
author = {Zaharia, Matei and Xin, Reynold S. and Wendell, Patrick and Das, Tathagata and Armbrust, Michael and Dave, Ankur and Meng, Xiangrui and Rosen, Josh and Venkataraman, Shivaram and Franklin, Michael J. and Ghodsi, Ali and Gonzalez, Joseph and Shenker, Scott and Stoica, Ion},
title = {Apache Spark: a unified engine for big data processing},
year = {2016},
issue_date = {November 2016},
publisher = {Association for Computing Machinery},
address = {New York, NY, USA},
volume = {59},
number = {11},
issn = {0001-0782},
url = {https://doi.org/10.1145/2934664},
doi = {10.1145/2934664},
journal = {Commun. ACM},
month = oct,
pages = {56–65},
numpages = {10}
}

@misc{SOCT,
  author = {Malachi Schram, Kishan Rajput, Armen Kasparian},
  title = {Scientific Optimization Control Toolkit (SOCT)},
  year = {2024},
  publisher = {GitHub},
  journal = {GitHub repository},
  howpublished = {\url{JeffersonLab/SciOptControlToolkit}},
}

@misc{towers2025gymnasiumstandardinterfacereinforcement,
      title={Gymnasium: A Standard Interface for Reinforcement Learning Environments}, 
      author={Mark Towers and Ariel Kwiatkowski and Jordan Terry and John U. Balis and Gianluca De Cola and Tristan Deleu and Manuel Goulão and Andreas Kallinteris and Markus Krimmel and Arjun KG and Rodrigo Perez-Vicente and Andrea Pierré and Sander Schulhoff and Jun Jet Tai and Hannah Tan and Omar G. Younis},
      year={2025},
      eprint={2407.17032},
      archivePrefix={arXiv},
      primaryClass={cs.LG},
      url={https://arxiv.org/abs/2407.17032}, 
}

@inproceedings{Kreps2011KafkaA,
  title={Kafka : a Distributed Messaging System for Log Processing},
  author={Jay Kreps},
  year={2011},
  url={https://api.semanticscholar.org/CorpusID:18534081}
}

@article{docker,
author = {Merkel, Dirk},
title = {Docker: lightweight Linux containers for consistent development and deployment},
year = {2014},
issue_date = {March 2014},
publisher = {Belltown Media},
address = {Houston, TX},
volume = {2014},
number = {239},
issn = {1075-3583},
journal = {Linux J.},
month = mar,
articleno = {2}
}

@software{influxdb,
  author       = {{InfluxData}},
  title        = {{InfluxDB}: Scalable Datastore for Metrics, Events, and Real-Time Analytics},
  url          = {https://github.com/influxdata/influxdb},
  year         = {2025},
  note         = {Accessed: 2026-03-06}
}

@software{grafana,
  author       = {{Grafana Labs}},
  title        = {{Grafana}: The Open and Composable Observability and Data Visualization Platform},
  url          = {https://github.com/grafana/grafana},
  year         = {2025},
  note         = {Accessed: 2026-03-06}
}

@article{td3,
  author       = {Scott Fujimoto and
                  Herke van Hoof and
                  David Meger},
  title        = {Addressing Function Approximation Error in Actor-Critic Methods},
  journal      = {CoRR},
  volume       = {abs/1802.09477},
  year         = {2018},
  url          = {http://arxiv.org/abs/1802.09477},
  eprinttype    = {arXiv},
  eprint       = {1802.09477},
  bibsource    = {dblp computer science bibliography, https://dblp.org}
}

@misc{Kasparian,
  title        = {Documentation for SMOCS},
  author       = {Kasparian, Armen},
  year         = {2026},
  howpublished = {\url{https://pages.jlab.org/datascience/smocs_docs/}},
  note         = {GitLab Documenation repository},
  url          = {https://pages.jlab.org/datascience/smocs_docs/}
}

\end{document}